\documentclass[12pt,a4paper,final]{iopart}

\usepackage{iopams}
\usepackage{graphicx}
\usepackage[breaklinks=true,colorlinks=true,linkcolor=blue,urlcolor=blue,citecolor=blue]{hyperref}

\usepackage{verbatim}
\usepackage[normalem]{ulem} 
\usepackage{upgreek}
\usepackage[disable]{todonotes}

\begin{document}

\title{Velocity asymmetry of Dzyaloshinskii domain walls in the creep and flow regimes}

\author{M.~Va\v{n}atka}
\address{CNRS, Institut N\'{e}el, 38042 Grenoble, France}
\address{Univ.~Grenoble Alpes, Institut N\'{e}el, 38042 Grenoble, France}
\address{Institute of Physical Engineering, Brno University of Technology, Technick\'{a} 2, 616 69 Brno, Czech Republic}

\author{J.-C.~Rojas-S\'{a}nchez}
\address{CNRS, Institut N\'{e}el, 38042 Grenoble, France}
\address{Univ.~Grenoble Alpes, Institut N\'{e}el, 38042 Grenoble, France}

\author{J.~Vogel}
\address{CNRS, Institut N\'{e}el, 38042 Grenoble, France}
\address{Univ.~Grenoble Alpes, Institut N\'{e}el, 38042 Grenoble, France}

\author{M.~Bonfim}
\address{Departamento de Engenharia El\'{e}trica, Universidade Federal do Paran\'{a}, Curitiba, Brazil}

\author{A.~Thiaville}
\address{Laboratoire de Physique des Solides, Univ. Paris-Sud, CNRS UMR 8502, 91405 Orsay, France}

\author[cor1]{S.~Pizzini}
\address{CNRS, Institut N\'{e}el, 38042 Grenoble, France}
\address{Univ.~Grenoble Alpes, Institut N\'{e}el, 38042 Grenoble, France}
\eads{\mailto{stefania.pizzini@neel.cnrs.fr}}

\begin{abstract}
We have carried out measurements of domain wall dynamics in a Pt/Co/GdO$_x(t)$ wedge sample with  perpendicular magnetic anisotropy. When driven by an easy-axis field $H_{z}$ in the presence of an in-plane field $H_{x}$, the domain wall expansion along $\pm x$  is anisotropic, as expected for samples presenting Dzyaloshinskii-Moriya interaction.
 In the creep regime, the sign and the value of the domain wall velocity asymmetry  changes along the wedge. We show that in our samples the domain wall speed \textit{vs.} $H_{x}$ curves in the creep regime cannot be explained simply in terms of the variation of the domain wall energy with $H_{x}$, as suggested by previous works. For this reason  the strength and the sign of the Dzyaloshinskii-Moriya interaction (DMI) cannot be extracted from these measurements. To obtain reliable information on the DMI strength using magnetic field-induced domain wall dynamics, measurements have been performed with high fields, bringing the DW close to the flow regime of propagation. In this case we find large values of DMI, coherent with those obtained from current-driven domain wall dynamics.

\end{abstract}


\vspace{2pc}
\todo[inline, author=Marek]{Add keywords}

Chiral magnetic textures such as Dzyaloshinskii domain walls (DDW) \cite{Thiaville2012}
and skyrmions \cite{Skyrme1960}  are attracting  attention because of their possible applications as information carriers in spintronics devices. DDW are N\'{e}el walls with a fixed chirality, stabilised, in non-centrosymmetric stacks, by the Dzyaloshinskii-Moriya interaction (DMI) \cite{Dzyaloshinskii1957,Moriya1960} present at the interface between a  magnetic layer and a heavy metal with large spin-orbit coupling. When driven by a Spin Hall effect related spin-orbit torque (SHE-SOT) \cite{LiuPRL2012,Haazen2013,Garello2013} DDW  in systems with perpendicular magnetic anisotropy (PMA) move with  large efficiency \cite{Miron2011,Ryu2013,Emori2013}. Also, it has been predicted that isolated skyrmions injected in nanotracks can be moved with very low current density and are moreover insensitive to defects \cite{Sampaio2013}.
Engineering materials with large  DMI has therefore become an important issue both for domain wall and skyrmion physics.

So far \textit{ab-initio} calculations of interfacial DMI are rare and concern perfect interfaces difficult to compare with the mixed interfaces found in  \textquotedblleft real\textquotedblright ~samples \cite{Freimuth2014,Yang2015}.
The information presently available on the DMI strengths relies on experimental work. A large input has been given by  Spin-polarised Scanning Tunneling Microscopy measurements  that show the presence of chiral magnetic textures or skyrmions in systems consisting of one monolayer of Fe (or Mn) on heavy metal substrates \cite{Bode2007,Ferriani2008,Meckler2009,Heinze2011} in ultra-high vacuum and at low  temperature. In the last few years, domain wall dynamics and nucleation measurements at room temperature have revealed the presence of DMI in less ordered, non centrosymmetric ultrathin magnetic layers with PMA, made by magnetron sputtering \cite{Ryu2013,Emori2013,Haazen2013,Pizzini2014}. More recently, Brillouin light scattering experiments have also highlighted the presence of DMI in similar PMA samples \cite{DiPRL2015, Belmeguenai2015}.

\begin{figure}[ht!]
\centering
\includegraphics[width=15cm]{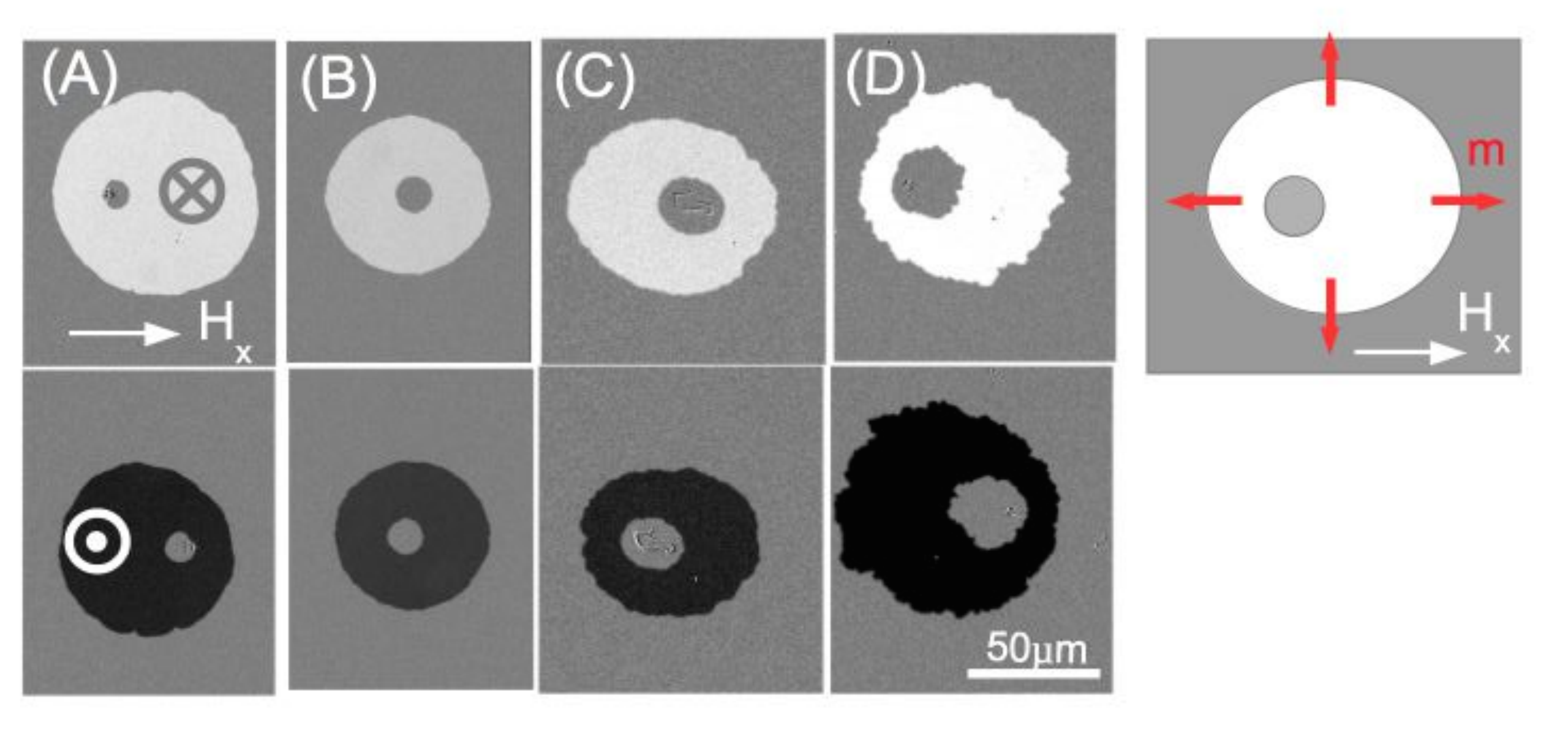}
\caption{\label{fig:images} Left: Expansion by DW propagation of an up (black contrast) and a down (white contrast) domain  in samples (A), (B), (C), (D). The Gd thickness increases from 2\,nm to 5\,nm going from (A) to (D).  The DW displacements are obtained by an $H_{z}$ field pulse with amplitude $\sim$10\,mT and duration $\sim$20-100\,ms  and a continuous in-plane field $H_{x}=+200$\,mT. Right: schematic view of a bubble domain expansion: the red arrows represent the equilibrium orientation of the magnetisation at the center of the DWs.}
\end{figure}

It has been shown recently that
when,  in a nanostrip or in a bubble domain, an easy-axis field $H_{z}$ drives the DW dynamics in the presence of an in-plane field $H_{x}$ (aligned along $+x$), the DW speed is different for up/down and down/up DDWs propagating along $\pm x$ \cite{Je2013,Hrabec2014,Jue2015}. This phenomenon is related to the symmetry  breaking introduced by the in-plane field. The  Dzyaloshinskii-Moriya interaction acts as a longitudinal chiral field
$H_{\mathrm{DMI}}=D/(\mu_{0}M_{s}\Delta)$ (where $D$ is the DMI strength, $M_{s}$ is the saturation magnetisation and $\Delta$ is the domain wall width parameter) localised on the domain walls, having  opposite directions for up/down and down/up DWs. Beyond a critical strength, the DMI forces the DW magnetisation in the N\'{e}el configuration  (see sketch in Figure~\ref{fig:images})  \cite{Thiaville2012}. Although the in-plane field does not drive the dynamics, it will respectively stabilise (\textit{vs.} destabilise) the DWs having their magnetisation $m$ parallel (\textit{vs.} antiparallel) to it. For a parallel (\textit{vs.} antiparallel)  alignment between $H_{x}$ and $m$ the DW speed increases (\textit{vs.} decreases) with respect to the $H_{x}$=0 case.
In the high speed (flow) regime, the speed increase (vs. decrease)  is mainly due to the widening (vs. narrowing) of the DW  with $H_{x}$ \cite{Jue2015}. In the low speed (thermally-activated or creep) regime, the speed dependence on $H_{x}$ has been related to the variation  of domain wall energy \cite{Je2013}.
 The DDW width (resp. DW energy) is expected  to have a minimum (resp. maximum) value when the applied in-plane field is equal and opposite to the stabilising $H_{\mathrm{DMI}}$ field i.e. when the DW acquires a Bloch form. In the two DW propagation regimes, this is the $H_{x}$ field for which the DW speed is predicted to exhibit a minimum.
 With these assumptions, $H_{x}$ could therefore be  a direct measure of the DMI energy density $D$, provided that the domain wall width parameter  $\Delta= \sqrt{A/K_{0}}$ ($K_{0}$ being the effective uniaxial anisotropy and $A$ the exchange constant) and  $M_{s}$ are known.

In the following, we will show that the DW speed~\textit{vs.}~in-plane field curves in the creep regime cannot in general be used to extract the strength and the sign of the DMI, as was done for Pt/Co/Pt samples \cite{Je2013,Hrabec2014}. Moreover, we find that the $v(H_{x})$ curves measured for the same sample in the thermally activated and in the flow regimes can have different trends.  Although the mechanism determining the exact trend of the velocity curves in the creep regime is not clear, we show that it cannot always  be described simply in terms of the variation  of DW energy with $H_{x}$.
Our measurements on Pt/Co/GdO$_x$ films suggest that modifications of the pinning barrier landscape upon application of the in-plane field also contribute to the trend of the  $v(H_{x})$ curves.

A Pt(5\,nm)/Co(1\,nm)/Gd($t$) stack with varying Gd thickness ($t=2-5\,$nm) was grown on a Si/SiO$_{2}$ substrate by magnetron sputtering in the shape of a wedge, and oxidised by O$_{2}$ plasma for 35 seconds. Consequently 2\,nm of Al were deposited on top of the stack to protect it from further oxidation.  The varying thickness of the Gd layer is at the origin of a gradient in the oxygen content at the Co/Gd interface, which varies the interfacial anisotropy \cite{Manchon2008a}.  All the samples present a well defined  PMA, with in-plane saturation fields varying between 1.6\,T (for 2\,nm Gd) and 0.6\,T (for 5\,nm Gd). Domain wall dynamics was studied at room temperature by wide-field Magneto-Optical Kerr microscopy, using a combination of easy-axis and in-plane magnetic fields. $H_{z}$ pulses of amplitude $\sim$10\,mT and duration   $\sim$20-100\,ms were obtained using a conventional, uncooled coil.  The $H_{z}$ pulses, driving the displacement of the DWs, were applied in the presence of a continuous in-plane field $H_{x}$, along $\pm
x$, which tunes the stability of the DDW internal structure. With such amplitudes of the $H_{z}$ field, DW speeds are of the order of some 0.1\,mm/s, the dynamics is  thermally activated and described by the so-called creep regime.

Starting from  (down or up) saturation,  a bubble domain was created by applying an up or a down $H_{z}$ pulse. The image of the domain was saved as a reference image. An $H_{z}$ pulse was then applied to enlarge the domain by DW propagation, and the new image was acquired. The difference between the two images gives the domain wall displacement that occurred during the field pulse. A black~(white) contrast in the images corresponds to the expansion of an up~(down) domain. The domain wall speed in a given direction can then be extracted from the ratio of the DW displacement and the pulse duration. DW displacements in the $\pm x$ directions were measured for a fixed value of the $H_{z}$ field, for several values of the in-plane field between -300~mT and +300~mT. In order to correct the residual $H_{z}$ component that may arise from a misalignment of the in-plane electromagnet, measurements were taken for both down and up domains.

Figure~\ref{fig:images} shows the differential images recorded in four positions of the wedge sample (called  samples (A) to (D) from now on) corresponding to increasing values of the Gd thickness (from 2~to~5\,nm) for $H_{z}$  field pulses of the order of $10$\,mT and an in-plane field of  $+200$\,mT. Without in-plane field,  the propagation of the DWs is isotropic and the domains are round. Similar to previously reported experiments, the $H_{x}$ field breaks the rotational symmetry and the propagation becomes asymmetric in the $\pm x$ directions. Note that the sign and the amplitude of the speed asymmetry depend on the sample composition. Indeed,  in sample  (A) the down/up DWs  move faster than the up/down DWs while in sample  (B) the asymmetry is practically vanishing, \textit{i.e.} up/down and down/up DWs move at the same speed. In sample  (C) the DW speed asymmetry reverses with respect to (A), \textit{i.e.} the up/down DWs move faster. Finally, in (D) the asymmetry found in (A) is recovered.

According to previous work \cite{Je2013,Hrabec2014}, the cancellation (resp. change of sign) of the  DW speed asymmetry may be attributed to a vanishing (resp. reversed) value of the DMI. This result is unexpected and  counter-intuitive. As one moves across the sample, from (A) to (D), the decreasing degree of oxygen content  modifies the composition of the Co/Gd interface, as shown experimentally by the changing PMA. However the sample presents a considerable  PMA  for the thinner Gd layers, which is an indication that the oxidation only concerns the top Co interface. Therefore the bottom Pt/Co interface, which is expected to provide the most important contribution to the DMI  \cite{Freimuth2014},  should not be affected by the varying Gd thickness. This is confirmed by X-ray reflectivity data.

\begin{figure}[ht!]
\centering
\includegraphics[width=15cm]{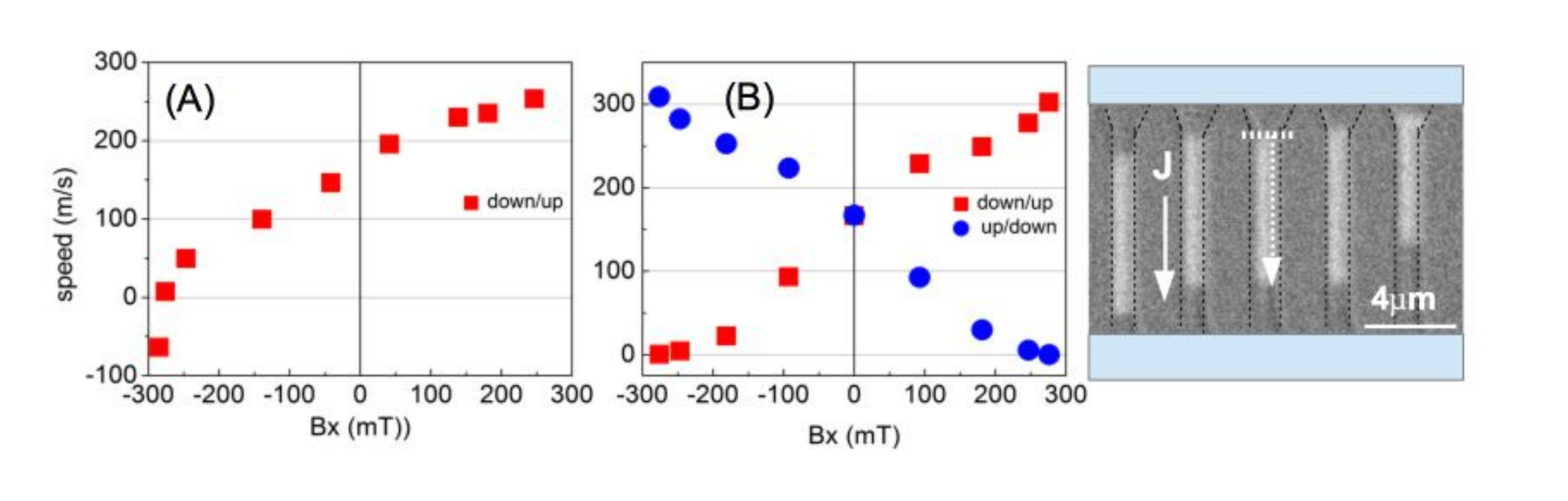}
\caption{\label{fig:speed_J} Left: Domain wall speed \textit{vs.} $H_{x}$ curves measured with constant current density $J=1.2 \times 10^{12}\,$A/m$^{2}$  for samples (A) and (B), for which a large and a vanishing field-induced domain wall speed asymmetry are found in the creep regime respectively. Right:  Differential Kerr image showing an example of the displacement of DWs in nanostrips.}
\end{figure}

In order to clarify the interpretation of the DW dynamics in the creep regime and to have an independent measurement of the sign of the DMI, we carried out current-induced DW dynamics measurements. For this purpose, the samples were patterned into $1\,\upmu$m wide strips by e-beam lithography and the DW dynamics was studied for a fixed value of the current-density $J=1.2 \times 10^{12}\,$A/m$^{2}$ and variable values of $H_{x}$. The results show that for all samples (note that sample  (C) could not be measured, due to deterioration during the patterning process)
the domain walls move in the same direction, opposite to the electron flow. Since in these systems the direction of the DW displacement is determined by the sign of the Spin-Hall angle in Pt (which is the same for samples (A) to (D)) and by the chirality of the DDW \cite{Thiaville2012,Ryu2013,Emori2013}, this results is a strong indication that the domain walls in all the samples  have the same chirality and therefore the sign of the DMI is sample independent.  The  results of the current-driven DW speed \textit{vs.} $H_{x}$ field curves for samples (A) and (B) are shown in Figure~\ref{fig:speed_J}.

The speed variation as a function of in-plane field $H_{x}$  is similar to that shown by other authors in strips of DMI materials \cite{Ryu2013,Emori2013,Ryu2014}.  In all the curves, the speed of the down/up DWs  increases for  positive $H_{x}$ fields and decreases for negative fields. The symmetric curve is found  for the up/down domain walls, as expected for chiral N\'{e}el walls. If we neglect the rotation of the magnetisation within the domains, the domain wall speed driven by the current $J$ \textit{via} the SHE-SOT can be expressed as \cite{Thiaville2012} :
\begin{equation}\label{DW_speed_vs_J}
v = \gamma_{0}\frac{\Delta}{\alpha}\frac{\pi}{2}\chi M_{s}\cos \psi
\end{equation}
where $\gamma_{0}$ is the gyromagnetic ratio, $\alpha$ is the damping parameter,  $\Delta$ is the domain wall width, $\psi$ is the angle of the DW magnetisation  with respect to the $x$-axis, and   $\chi=\hbar \theta_{H} J / (2e\mu_{0} M_{s}^{2} t)$ where $\theta_{H}$ is the Spin Hall angle and $t$ the magnetic layer thickness. It can be shown (see Suppl. Information) that for our samples the variation of $\cos \psi$ with $H_{x}$ is negligible except around  $H_{x}=-H_{\mathrm{DMI}}$  where it changes sign, so that the  $v(H_{x})$ shape is mainly determined by the modification of the domain wall width with $H_{x}$.
Since the DW width increases for an $H_{x}$  field parallel to the DW magnetisation, our measurement show that down/up DWs have their magnetisation parallel to the $+x$ direction and therefore that the DWs in the Pt/Co/GdO$_x$ samples have left-handed chirality, like in Pt/Co/AlO$_x$ \cite{Pizzini2014,Jue2015}. This is not surprising, as we expect that the DMI interaction is mainly located at the Pt/Co interface.

The velocity of the down/up DW in sample (A) changes direction under the effect of  a negative in-plane field $H_{x}\approx -280$\,mT; this is associated with the switching of the DW chirality when the negative $H_{x}$ field exceeds the local chiral $H_{\mathrm{DMI}}$ field.  This in-plane field value is therefore a measure of $H_{\mathrm{DMI}}$. Note that in sample (B) the switching of the DW velocity is hindered by the larger DW pinning  \cite{Ryu2013,Ryu2014}.

The constant sign of the DMI for all the samples - assessed by the constant direction of current-driven DW motion at zero $H_{x}$ field  - in contrast with the different DW velocity asymmetries observed for the different samples in Figure~\ref{fig:images},  sheds  doubts on the possibility to deduce the sign of the DMI from the domain expansion images in the creep regime. In order to clarify  the interpretation of the field-induced measurements, we measured the DW speeds as a function of  $H_{x}$ field for the bubble domains shown in Figure~\ref{fig:images}.

\begin{figure}[ht!]
\centering
\includegraphics[width=15cm]{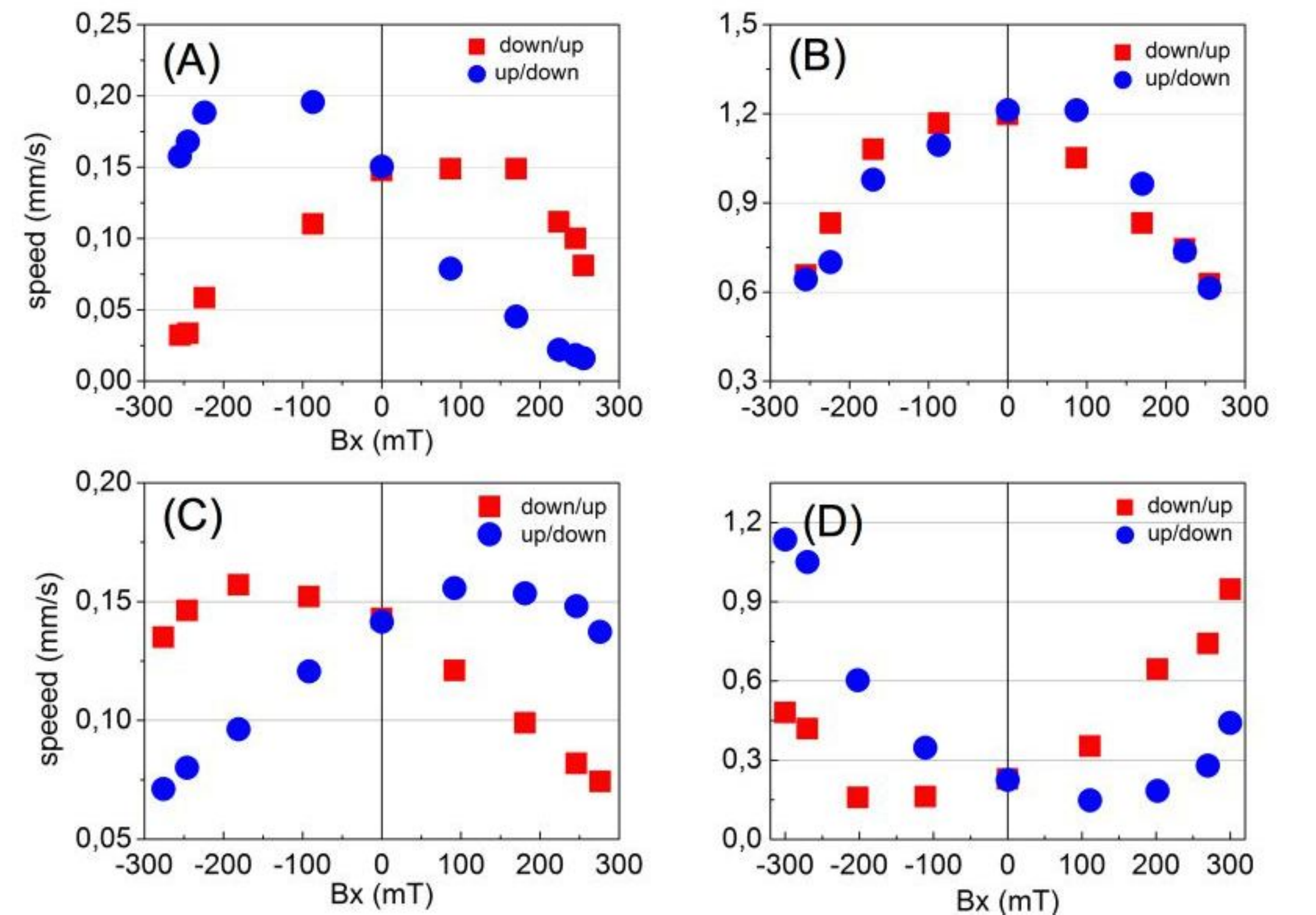}
\caption{\label{fig:speed_Hcreep} Domain wall speed \textit{vs.} $H_{x}$ field measured in the thermally activated regime for bubble domains in Pt/Co/GdO$_x$ samples (A) to (D), for the DW propagating along the $x$-axis direction.  }
\end{figure}

The velocity curves are shown in Figure~\ref{fig:speed_Hcreep} for
the two domain walls propagating along the $x$-axis  and having their magnetisation either parallel or antiparallel to the $H_{x}$ field.  The up/down and the down/up DWs exhibit the same behaviour for opposite $H_{x}$ fields, as expected for chiral N\'{e}el walls.  The curves for sample (D) - corresponding to the thicker Gd layer - present the main features found by other authors for DDWs in Pt/Co/Pt films \cite{Je2013,Hrabec2014}.  The speed of the down/up DW increases for a positive in-plane field, and for negative fields it decreases down to a minimum value between -100~mT and -200~mT, where the velocity starts increasing again. On the other hand, the curves measured for samples (A-C) strongly deviate from the expected behaviour, showing in particular a maximum rather than a minimum in the DW speed.

In the thermally activated regime, the DW velocity is given by:
\begin{equation}\label{speed}
v(H_{z)}=v_{o}\exp(-\eta H_{z}^{-\mu})
\end{equation}
where $v_{0}$ is the characteristic speed, $\mu$=1/4 is the creep scaling exponent and $\eta=U_{c} H_{\mathrm{crit}}^{\mu}/k_{B}T$ where $U_{c}$ is an energy scaling constant and $H_{\mathrm{crit}}$ the critical magnetic field \cite{Lemerle1998,Kim2009}.
Following Ref. \cite{Kim2009}, $U_{c}$ is related to $\xi$ (the correlation length of the pinning potential) and to the Larkin length $L_{c}=(\sigma_{\mathrm{DW}}^{2}t^{2} \xi^{2}/\gamma)^{1/3}$ (the characteristic length of rigid microscopic DW segments) and $H_{\mathrm{crit}}=\sigma_{\mathrm{DW}}\xi/M_{s}L_{c}^{2}$ where $\sigma_{\mathrm{DW}}$ is the DW energy and $\gamma$ is the pinning strength  of
the disorder. By assuming that neither $\xi$  nor $\gamma$ are modified by $H_{x}$,   Je \textit{et al.} \cite{Je2013} conclude that the shape of $v(H_{x})$ is solely due to the in-plane field dependence of the DW energy. According to \cite{Pizzini2014}, the energy of a DDW, taking into account the modification of the DW profile with $H_x$   reads:
\begin{equation}\label{full-DW-profile}
\sigma=\sigma_{00}[\sqrt{1-h^{2}}+(h+\frac{2}{\pi}\frac{D}{D_{c0}}) (\arcsin h \mp \pi/2)].
\end{equation}
where $\sigma_{00}=4\sqrt{AK_{0}}$ is the DW energy at rest, $D_{c0} = 4 \sqrt{A K_0}/\pi \equiv \sigma_{00}/\pi$ gives the onset of
magnetisation cycloids, $h= H_x / H_{K0}$ and the $\mp$ signs refer to the DW having its magnetisation parallel/antiparallel to the $H_x$  field. The energies of the DW favoured/unfavoured by the in-plane field  are the same
when $h=-(2/\pi) (D/D_{c0})$ or when $H_{x}= -D/(\mu_{0}M_{s}\Delta)$ = -$H_{\mathrm{DMI}}$. This is the in-plane field for which the DW energy is maximum. From equation \ref{speed} it then follows that the DW velocity should exhibit a minimum for $H_{x}=-H{_\mathrm{DMI}}$. This is indeed observed for sample (D). Note that the left-handed DW chirality deduced from the measurement agrees with the results of the current-induced measurements.

The speed \textit{vs.} $H_{x}$ curves obtained for samples  (A-C) show a different behaviour. For sample (A) the speed asymmetry is the same as for sample (D), but the velocity of the down/up DW increases for negative fields and decreases up to a critical field for positive fields. In sample (B) the speed asymmetry practically disappears and the speeds of the up/down and down/up DWs continuously decrease with both positive and negative $H_{x}$ fields.  In sample (C) the asymmetry is switched for  down/up and up/down DW, with respect to sample (D).  Therefore in these three samples the  $v(H_{x})$ curves do not follow the variation of the DW energy. Curves deviating from the expected behaviour have also been recently reported in the literature  \cite{Lavrijsen2015}.

Note that the \textquotedblleft anomalous\textquotedblright~curves are found in particular for samples (B) and (C),  for which the sign of the speed asymmetry would suggest that the value of $D$ is either vanishing (for (B)) or opposite (for (C)) to the  one of sample (D). This indicates that in the creep regime extreme care should be taken when extracting information on the DMI sign and amplitude simply on the basis of  the asymmetry (or lack of asymmetry) of the Kerr microscopy differential images. Before assessing about $D$, the full speed\textit{ vs.} $H_{x}$  curves should be examined and compared with the curves predicted by the existing  theoretical models.

\begin{figure}[hb!]
\centering
\includegraphics[width=15cm]{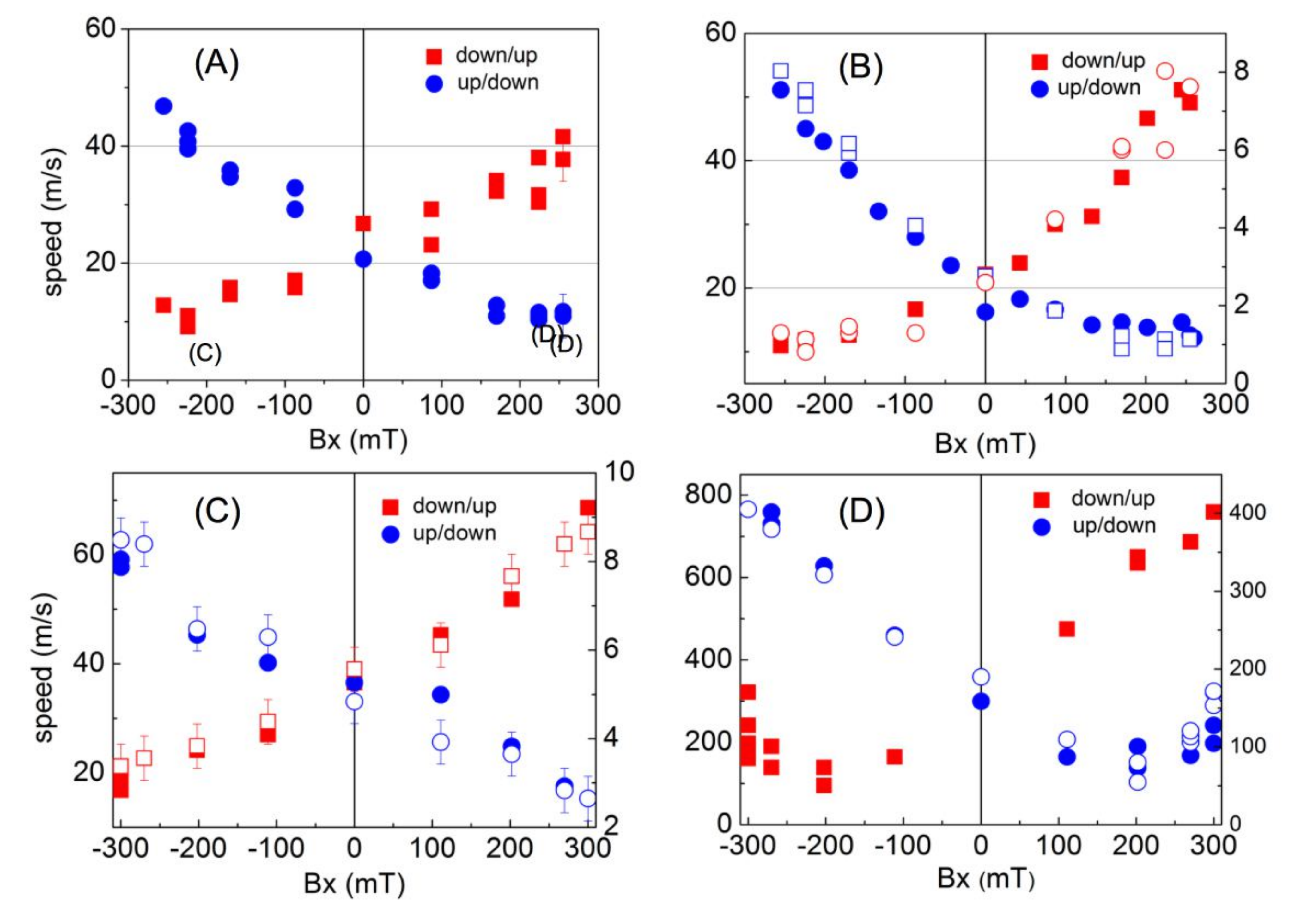}
\caption{\label{fig:speed_H_flow} Domain wall speed \textit{vs.} $H_{x}$ field measured for bubble domains in Pt/Co/GdO$_x$ for samples (A) to (D) for 20ns-long $H_{z}$  field pulses varying between 70mT and 200mT.   For samples (B) to (D)  the curves were  measured for two $H_{z}$ field values (empty symbols correspond to the scale to the right). The trends of the normalised speed curves are the same for each field value. For sample (D) the speeds are larger, as the depinning of the DWs occurs for lower fields. }
\end{figure}

In order to verify the role of the DW pinning on the speed \textit{vs.} $H_{x}$ field,  we have repeated the field-dependent measurements for larger values of the $H_{z}$ fields, bringing the domain wall velocities to a regime ($\gg$ 1\,m/s) where the propagation is much less sensitive to the pinning generated by local variations of the anisotropy field. Pulsed $H_{z}$ fields up to 200\,mT and duration down to 20\,ns were obtained using a $50\,\upmu$m wide microcoil coupled to a fast current pulse generator \cite{Mackay2000}.
The results reported in Figure~\ref{fig:speed_H_flow} for samples (A) to (D) show that in these conditions the speed \textit{vs.} $H_{x}$ curves all acquire the trend expected for chiral N\'{e}el walls in the flow regime \cite{Jue2015}.

In the high field regime, the stationary DW velocity is given by $v=\gamma_{0}\Delta_{T} H_{z}/\alpha m_{z0}$, where
$\Delta_{T}$ is the Thiele domain wall width \cite{Thiele1974} and $m_{z0}$ is the easy-axis magnetisation within the domains. The speed variation with $H_{x}$ is mainly related to the
modification of the Thiele DW width with the in-plane field (see Suppl. Information). In all the samples, the down/up  DWs propagate faster than the up/down DWs for positive $H_{x}$  fields, confirming once again that the DW chirality is the same (left handed) in agreement with the current-induced measurements and the field-induced (creep) measurements for sample (D).  For samples (A) to (C) the DW speed of the down/up DWs  decrease down to the largest available negative $H_{x}$ field, with a saturation but not a clear minimum in the DW speed. This suggests that $H_{\mathrm{DMI}}$ in these samples is of the order or more than $+300$\,mT.

For sample (D), where the PMA (and therefore the $H_{\mathrm{DMI}}$ field) is reduced, the down/up DWs exhibit minimum speed for $H_{x}=\approx-180$\,mT, a value close to that found for the same sample in the creep regime. By taking $M_{s}=1 \times 10^{6}$\,A/m (measured by VSM-SQUID),   $\mu_{0}H_{K}=0.7$\,T (measured by EHE) and $A=2.2 \times 10^{-11}$\,J/m \cite{Metaxas2007} the expression $H_{\mathrm{DMI}}=D/(\mu_{0}\Delta~M_{s})$  gives rise to a value of $D=1.27$\,mJ/m$^{2}$, with $\Delta=7.1$\,nm. Taking into account the larger thickness of the Co layer (1\,nm) in our Pt/Co/GdO$_x$ samples, this value scales reasonably well with the $D=2$\,mJ/m$^{2}$ value found for Pt/Co(0.6\,nm)/AlO$_x$ \cite{Pizzini2014,Jue2015}. For samples (A) to (C) it is difficult to obtain a precise value of $D$ from the field-dependent measurements, where the minimum speed is not well defined. The value of $D$  for sample (A) may be derived from  the in-plane field for which the DW chirality switches when driven by spin-polarised current ($H_{x}=\approx-280$\,mT). Using the values of $M_{s}$ and $A$ used for sample (D), and the measured in plane saturation field $\mu_{0}H_{K}=1.6$\,T
 giving rise to $\Delta=4.7$\,nm, we obtain a value of $D=1.31$\,mJ/m$^{2}$. We estimate that the uncertainty associated to the value of the exchange parameter A, together with the error associated to the definition of the $H_{x}$ field where the DW velocity is minimum, allow the determination of $D$ with a precision not better than $\pm0.2\,$mJ/m$^{2}$.
 The similar $D$ values found for the two samples
  indicate that the DMI strength is homogeneous along the wedge sample and that it is mainly arising from the Pt/Co interface.

As a consequence,  the \textquotedblleft anomalous\textquotedblright~$v(H_{x}$) curves in the creep regime do not bear any information about the sign and strength of the DMI.  In  Figure~\ref{fig:speed_Hcreep},  the value for which the speed is maximum in samples (A) and (C) is not related to the $D$  value, and the absence of speed asymmetry for sample (B) is not a signature of a vanishing $D$.  Since the anomalous behaviour of the $v(H_{x}$) curves is observed only in the creep regime and for samples (A) to (C), we conclude that this feature  may be related to  modifications of the domain wall pinning with $H_{x}$, which depends on the details of the Co/Gd interface. Since the measurements were taken with ms-long pulses in the creep regime and with ns-long pulses in the flow regime, the effect of the pulse length on the DW pinning may also play a role.

\begin{figure}[ht!]
\centering
\includegraphics[width=9cm]{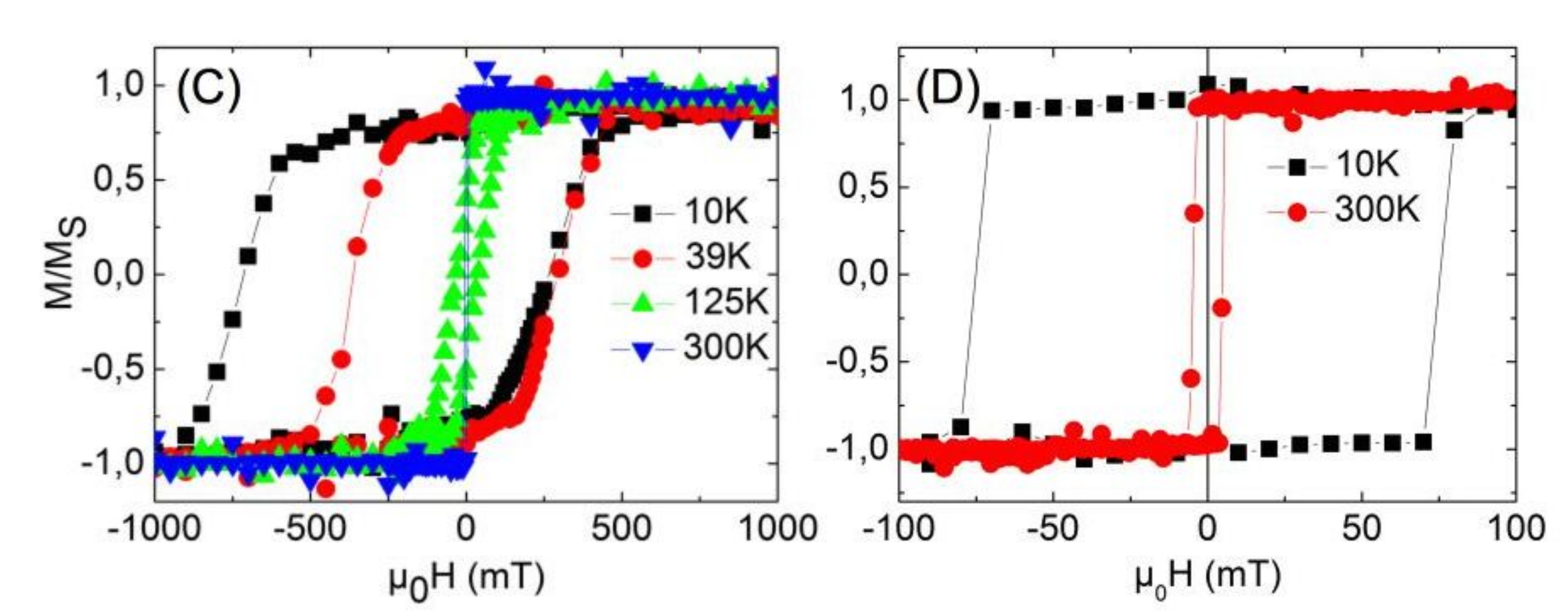}
\caption{\label{fig:VSM} VSM-SQUID measurements carried out from 10\,K to 300\,K for an out-of-plane field up to 1\,T.  Left:  in sample (C), the shift of the cycle at low temperature is an indication of the presence of CoO at the top Co/Gd interface. Right:  in sample (D), the cycle does not exhibit a shift, sign of the absence of relevant oxidation. Note the factor 10 difference in the field scale.}
\end{figure}

Some information on the nature of the top interface, as the presence or not of CoO, can be obtained from the temperature dependence of magnetic hysteresis loops. We have carried out magnetisation measurements with variable temperature between 10\,K and 300\,K in a VSM-SQUID of Quantum Design (Figure~\ref{fig:VSM}). For  sample (C) a change of the hysteresis loops, which are square with 100\% remanence at 300\,K, is observed around 225\,K, where they become partly tilted and the remanence decreases to about 60\%. This indicates a decrease of the PMA. Upon decreasing the temperature further, the coercivity increases strongly and below 70\,K a shift of the hysteresis loop to negative fields develops. Both observations can be attributed to the presence of an ultrathin layer of CoO at the Co/Gd interface, which becomes antiferromagnetic around 225\,K with a blocking temperature around 70\,K. For sample (D), the only one presenting
\textquotedblleft expected\textquotedblright~$v(H_{x}$) curves in the creep regime,  the cycles do not exhibit any exchange bias indicating that no CoO is formed at the top Co interface.

The \textquotedblleft anomalous\textquotedblright~behaviour of the $v(H_{x}$) curves in the creep regime seems therefore to be related to the presence of Co oxide at the top Co interface, and the details of the curves to the different degree of oxidation. Although the CoO is paramagnetic at the room temperature, it exhibits a magnetic susceptibility in the $x$-direction \cite{Ambrose1996}. We speculate that the CoO magnetic moments induced in the $x$ direction by the in-plane field  may act as an extra pinning potential acting on the DWs.  Since the magnetic susceptibility may depend on the CoO thickness, this could explain why different samples exhibit maximum velocity for different $H_{x}$ fields.  The description of the creep law simply in terms of the variation of the DW energy may not be general, as the pinning potential landscape may be strongly affected by the in-plane field.

In conclusion, we have shown that in Pt/Co/GdO$_x$ samples with different oxidation degrees of the Co/Gd interface the dependence of the DW velocity as a function of the in-plane field  cannot be interpreted within the creep law relating the DW speed changes exclusively to  the DW energy variations. Therefore in these samples, the $v(H_{x}$) curves fail to give information about the sign and the strength of the DM interaction. We have correlated the failure of the proposed creep law with the modification of the pinning potential landscape induced by the in-plane field.
When by applying strong  and ultrashort out-of-plane field pulses we change the dynamic regime of the DW propagation, the $v(H_{x}$) curves indicate that the chirality of the DDW is left-handed, and $D$ is of the order of $1.3$\,mJ/m$^{2}$ for $1$\,nm Co.

This work was supported by the Agence Nationale de la Recherche, project ANR 11 BS10 008 ESPERADO.
SP acknowledges the support of E. Wagner and of the staff of the Nanofab facility in Institut N\'{e}el.

\newpage
\textbf{SUPPLEMENTAL MATERIAL}

\section{MICROMAGNETIC SIMULATIONS}

Micromagnetic simulations of domain wall (DW) velocities driven either by an out-of-plane magnetic field or by an electrical current in a nanostrip were performed using a full 1D micromagnetic model, as introduced in \cite{Thiaville2012}. The unknown parameter is the full $\vec{m} (\vec{x}, t)$ profile. The demagnetizing field is evaluated by direct summation for a given  nanostrip width (500 nm). In our case, as the sample thickness and the DW width are much smaller than the typical strip width, the wall may be considered as infinitely long. The computed value of the demagnetizing field is averaged over the sample thickness and over the strip width. In practice, only the $x$ component of the demagnetizing field is evaluated; the $y$ component that is transverse to the strip is negligibly small and the $z$ component is approximated by a local value with unity demagnetizing factor. For the evaluation of DW dynamics, the finite box containing the DW is shifted along the strip so that the DW is always kept in its center.
For field-driven dynamics, the case of bubble domains in the continuous layers
 is too complex to be treated specifically; the DW propagation along a given direction has been assimilated to that of a domain wall within a strip aligned along that particular direction.
Having obtained the complete profile of magnetization in the domains and across the DW, all quantities of interest can be evaluated. These contain:

(i) the Thiele DW width \cite{Thiele1974}, defined as:
\begin{equation}\label{Thiele}
\frac{2}{\Delta_{T}}=\frac{1}{S}\int(\frac{\partial{\vec{m}}}{\partial{x}})^{2}d^{3}r
\end{equation}
where S is the cross-section area of the nanostrip, oriented along the $x$ direction.

(ii) the angle $\Phi$ of the DW magnetic moment which controls the force exerted by the Spin Hall Effect (SHE) on the DW in the case of current-driven DW motion. In the presence of an in-plane field and SHE the domain magnetization rotates so that the straightforward evaluation of $\cos\Phi$ fails. To remedy this, we evaluate by integration the SHE force on the DW. This is proportional to $\int(\vec{m}\times \partial_{x}\vec{m})_{y}$, with a value of $\pi\cos\Phi$ in the simple case, $\pi$ standing for the angle between the two domain magnetizations. Thus, to compute $\Phi$, this integral is numerically evaluated, and then divided by the angle between the domain magnetizations.

The micromagnetic parameters chosen for the simulations are $M_{s}$ = 1000 \,kA/m, A = 22 \,pJ/m,
$K_{u}$  = 0.87 and 1.17 \,MJ/m$^{3}$ where $K_{0} = K_{u}-\mu_{0}M^{2}_{s}/ 2 $ is the effective anisotropy including the perpendicular demagnetizing field effect in the local approximation. For the  strength of the Dzyaloshinskii-Moriya interaction we used the value $D$ = 1.8 \,mJ/m$^{2}$, which is slightly larger than the one evaluated for our samples.
For the magnetization dynamics, the gyromagnetic ratio of the free electron $\gamma_{0} = 2.21 \times 10^{5}$ \,m/(A$ \cdot s)$ and the damping factor $\alpha$ = 0.5 extracted from DW dynamics experiments were used.

The aim of these simulations is to explain qualitatively  the observed trends in the speed \textit{vs}. $H_{x}$ curves, in particular for samples with different magnetic anisotropy values.

\subsection{Current-driven dynamics}
\begin{figure}[ht!]
\includegraphics[width=16cm]{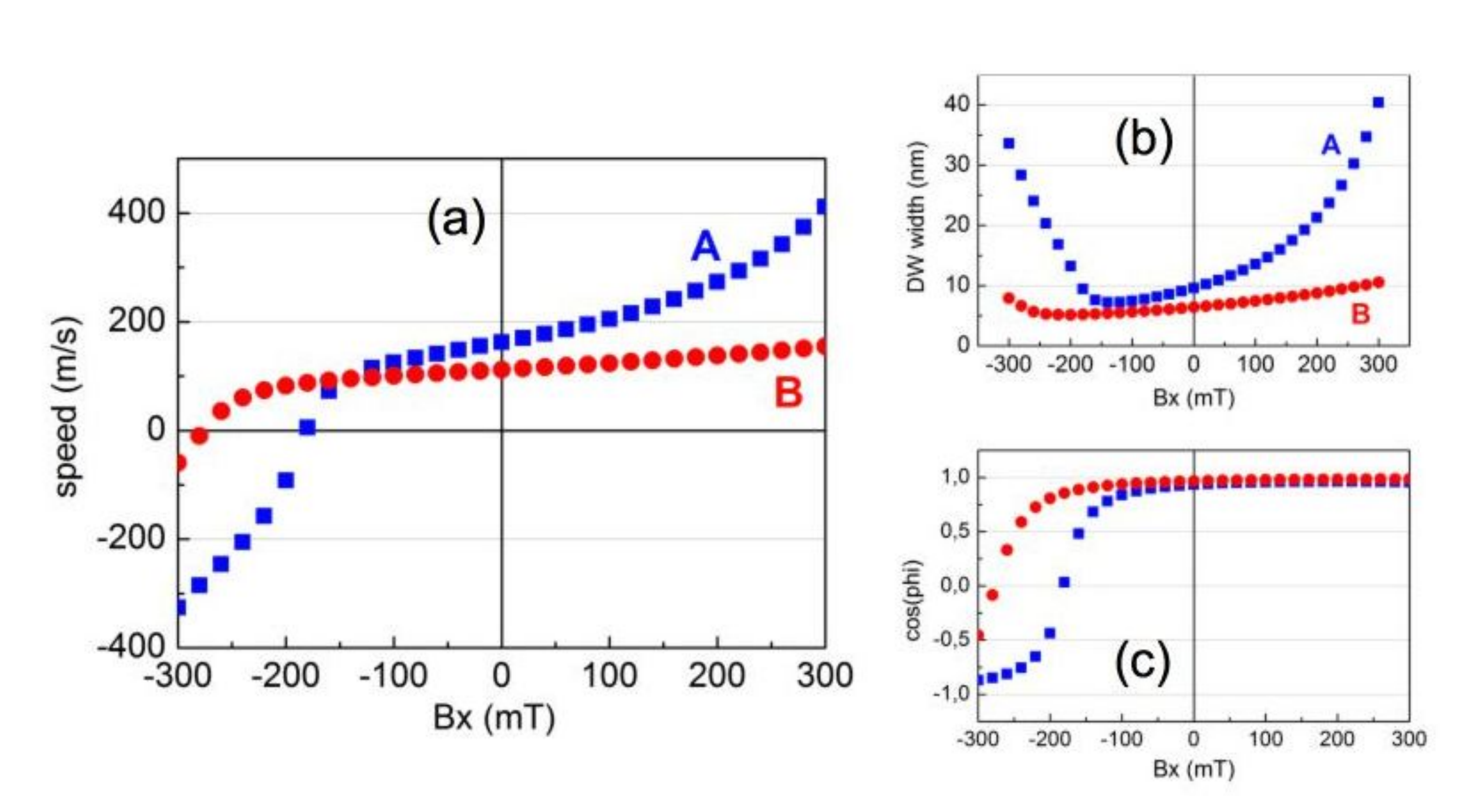}
\label{fig:Suppl-Figure1-speed-vs-Bx}\textbf{Figure S1} : (a) Current-driven domain wall speed \textit{vs.} $\mu_{0}H_{x}$ calculated for a fixed value of the current density $J=1\times 10^{12}$ \,A/m$^{2}$ and two values of the anisotropy energy  $K_{u}$ = 0.87 \,MJ/m$^{3}$ (curve A) and 1.17 \,MJ/m$^{3}$ (curve B); (b) variation of Thiele DW width with $\mu_{0}H_{x}$ ; (c) variation of $\cos\Phi$ with $\mu_{0}H_{x}$.
\end{figure}

The simulations were carried out for a fixed value of the current density  $J=1\times 10^{12}$ \,A/m$^{2}$.
Figure~\ref{fig:Suppl-Figure1-speed-vs-Bx} (a) shows the variation of the domain wall speed as a function of longitudinal in-plane field $H_{x}$, for two values of the out-of-plane-anisotropy  $K_{u}$  = 0.87 and 1.17 \,MJ/m$^{3}$ (A and B). The figure shows that the DW speed changes more rapidly for the low anisotropy value, in agreement with the experiments. For strong enough negative $H_{x}$ field (i.e. antiparallel to the DW magnetisation direction) the domain wall velocity changes sign, due to the reversal of the domain wall magnetisation. The reversal field ($H_{x}=-H_{DMI} =-D/(\mu_{0}\Delta M_{s})$) is smaller for the smaller anisotropy value, as the domain wall width is larger.  As seen in the main text, the detailed shape of the speed vs. $H_{x}$ curve depends both on the variation of the Thiele domain wall width $\Delta_{T}$ and on the value of $\cos\Phi$, where $\Phi$ is the angle that the DW magnetisation forms with the $x$-axis along  which the in-plane field is applied. Figures~\ref{fig:Suppl-Figure1-speed-vs-Bx} (b-c) illustrate the simulated values of $\Delta_{T}$ and $\cos\Phi$. For positive $H_{x}$ values the speed curve variations are only related to the variation of $\Delta_{T}$, which is larger for the smaller anisotropy. For negative fields, the change of the $\cos\Phi$ sign (reversal of the DW magnetisation direction) is at the origin of the reversal of the DW speed.

\subsection{Field-driven dynamics in creep and flow regimes}
\begin{figure}[ht!]
\includegraphics[width=16cm]{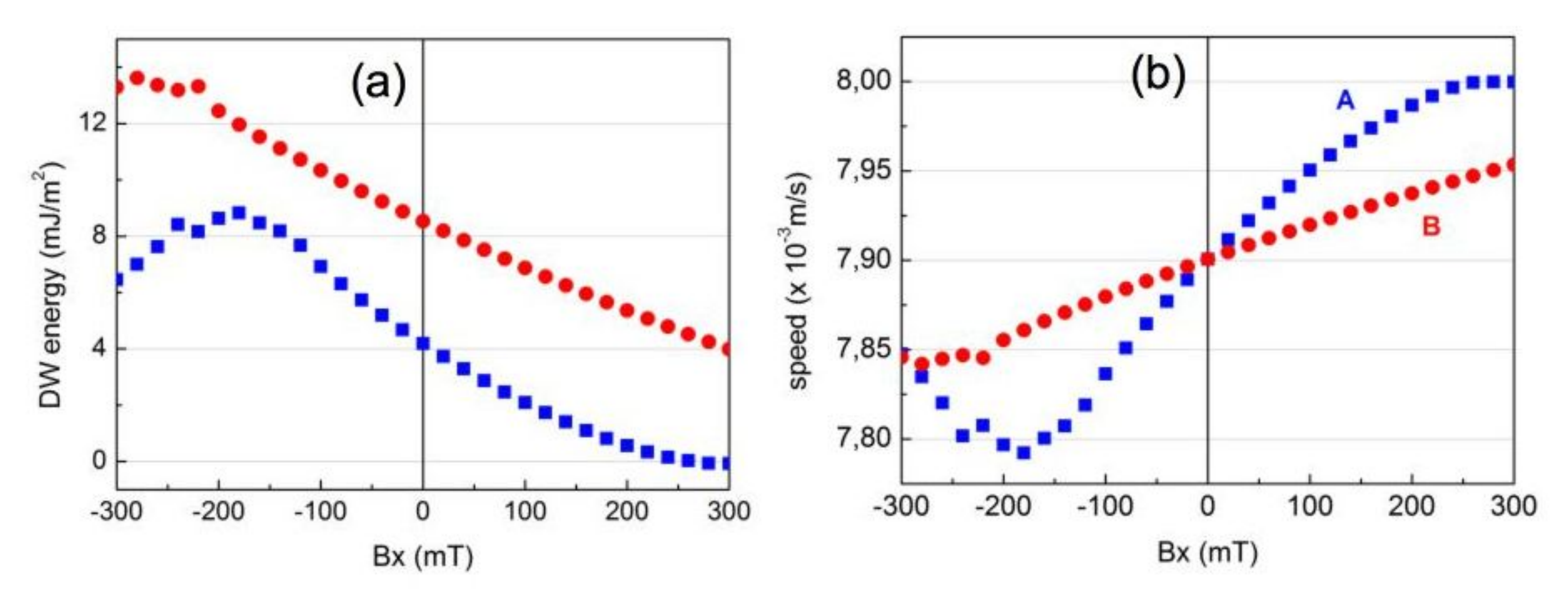}
\label{fig:Suppl-Figure2-DWenergy-vs-Bx}\textbf{Figure S2} : (a) Domain wall energy \textit{vs.} $B_{x}$ simulated for a fixed $B_{x}$ value of 50mT.  (b) variation of DW domain wall speed \textit{vs.} $\mu_{0}H_{x}$ assuming a $v_{0}$ value of $8\times10^{-3}$ \,m/s and $\eta=\sigma(H_{x})/\sigma(0)$ in equation (2) of the main text. (A) and (B) refer to the two anisotropy values.
\end{figure}

The field-driven domain wall dynamics under in-plane field is dependent on the working regime (creep or flow). The dynamics of chiral N\'{e}el domain walls in the presence of an in-plane field has been studied  in the creep regime \cite{Je2012,Hrabec2013} and in the flow regime \cite{Jue2015} . It has been shown that the $v(H_{x})$ curves can be explained in terms of the variation of the DW energy with the $H_{x}$ field.
Figure~\ref{fig:Suppl-Figure2-DWenergy-vs-Bx}(a) shows the variation of the domain wall energy \textit{vs.} $H_{x}$ for a sample having the magnetic parameters given above and $\mu_{0}H_{z}$=50 \,mT
. The DW energy is maximum for the
$H_{x}=-H_{DMI}$, where the DW has the Bloch form. This field is of course dependent of the anisotropy value.

The expected $H_{x}$ dependence of the DW speed in the creep regime is shown in  Figure~\ref{fig:Suppl-Figure2-DWenergy-vs-Bx}(b) for the two anisotropy values. The curves present a minimum for $H_{x}=-H_{DMI}$  and have a symmetric behaviour on either sides of this field. The behaviour measured for sample (D) is in qualitative agreement with these curves. That found for samples (A) to (C) is in strong disagreement, showing that the $v(H_{x})$ curves are not simply related to the change of the DW energy with $H_{x}$.

\begin{figure}[ht!]
\includegraphics[width=16cm]{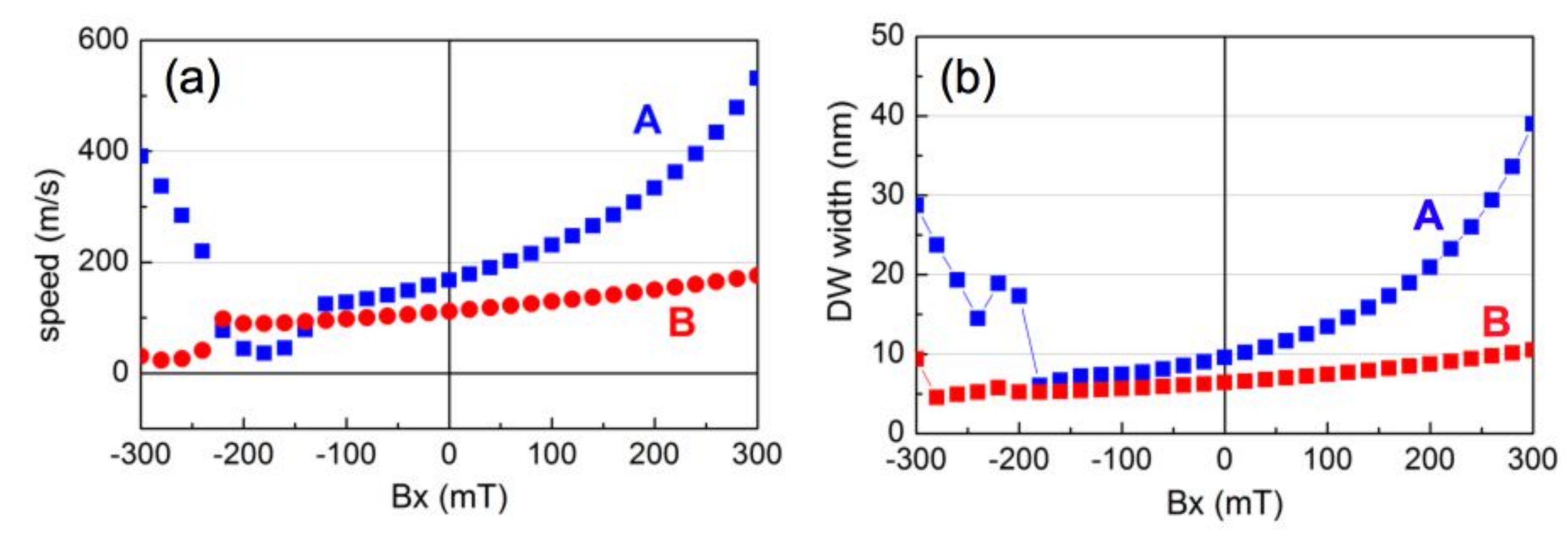}
\label{fig:Suppl-Figure3-DWspeed-vs-Bx-flow}\textbf{Figure S3} : (a) Domain wall speed \textit{vs.} $\mu_{0}H_{x}$ in the flow regime simulated using $\mu_{0}H_{x}$ = 50\,mT and the two anisotropy values given above; (b) the variation of the Thiele DW width.
\end{figure}

If we neglect the tilt of the magnetisation in the domains, in the flow regime the stationary DW velocity is given by $v=\gamma_{0}\Delta_{T} H_{z}/\alpha$ and the speed variation with $H_{x}$ is expected to be related to the modification of $\Delta_{T}$ with the in-plane field. The two curves simulated for a $\mu_{0}H_{x}$ = 50 \,mT are shown in Figure~\ref{fig:Suppl-Figure3-DWspeed-vs-Bx-flow}. As  in the case of current-driven dynamics, the Thiele DW width increases (decreases) for positive (negative) in-plane field, and in a larger extent for the low anisotropy value (A). The dip in the DW speed in the vicinity of  $H_{x}=-H_{DMI}$, corresponding to the precessional regime of the DW, is not observed in the experimental data. Although this discrepancy is still the object of our studies, we believe that the local changes of the DMI strength over the region swept by the DW during the measurements may contribute to the smoothening of the discontinuity associated to the precessional regime.

\end{document}